\documentclass[a4paper,11pt,oneside]{article}

\usepackage[english]{babel}
\usepackage[T1]{fontenc}
\usepackage[utf8]{inputenc}

\usepackage[pdftex,unicode]{hyperref}
\hypersetup{pdftitle=The Impact of Operating Environment on Efficiency of Public Libraries}
\hypersetup{pdfauthor=Vladimír Holý}

\usepackage[margin=60pt]{geometry}
\usepackage{amsmath}
\usepackage{amssymb}
\usepackage{bm}

\usepackage[group-minimum-digits=3]{siunitx}

\usepackage{enumitem}

\usepackage{graphicx}
\usepackage{multirow}
\usepackage{longtable}
\usepackage{booktabs}

\usepackage[authoryear]{natbib}

\begin{document}

\begin{center}
{\Large \bfseries The Impact of Operating Environment on Efficiency of Public Libraries}
\end{center}

\begin{center}
{\bfseries Vladimír Holý} \\
University of Economics, Prague \\
Winston Churchill Square 4, 130 67 Prague 3, Czech Republic \\
\href{mailto:vladimir.holy@vse.cz}{vladimir.holy@vse.cz}
\end{center}

\begin{center}
{\itshape \today}
\end{center}

\noindent
\textbf{Abstract:}
Analysis of technical efficiency is an important tool in management of public libraries. We assess the efficiency of \num{4660} public municipal libraries in the Czech Republic in the year 2017. For this purpose, we utilize data envelopment analysis (DEA) based on the Chebyshev distance. We consider total expenditures, employees and book collection as inputs with registrations, book circulation, event attendance and collection additions as outputs. We pay special attention to the operating environment and find that the efficiency scores significantly depend on the population of the municipality and its distance to the municipality with extended powers. To remove the effect of the operating environment, we perform DEA separately for categories based on decision tree analysis as well as categories designed by an expert.
\\

\\

\\

\section{Introduction}
\label{sec:intro}

On June 22, 1919 in the former Czechoslovakia, a law was passed introducing the obligation to establish a library in each municipality. One hundred years have passed and the Czech Republic with Slovakia have one of the densest networks of public libraries in the world. There was one public library for every \num{1983} citizens in the Czech Republic in 2017. Since June 29, 2001, Czech libraries are governed by the Law No. 257/2001 Coll. on Libraries and Terms of Operating Public Library and Information Services (Library Act). The size of public libraries ranges from small village libraries affiliated with the municipal offices to large library networks in major cities. The role of libraries has also changed over decades. The original purpose of collecting and lending books and periodicals has been supplemented by managing audio-visual materials, providing internet access, offering professional assistance, organizing community meetings and hosting cultural events. Such a large amount of diverse libraries with a wide range of activities requires well-advised management and careful allocation of public resources.

In this study, we analyze technical efficiency of Czech public libraries established by municipalities. We follow the data envelopment analysis (DEA) approach pioneered by \cite{Charnes1978} and \cite{Banker1984}. DEA is a non-parametric method measuring how efficiently can decision making units (DMU) transform a set of inputs to a set of outputs. We utilize the Chebyshev distance DEA model with variable returns to scale recently proposed by \cite{Hladik2019}. This model is based on the robust optimization viewpoint and has many desirable properties -- super-efficiency, comparability of efficiency scores across different analyzes, inclusion of zero inputs and outputs, units invariance, order of rankings identical to the classical approach and straightforward interpretability.

There are many studies in the literature assessing technical efficiency of libraries. We select input and output variables consistently with the literature. Specifically, we consider total expenditures, employees and book collection as inputs with registrations, book circulation, events attendance and collection additions as outputs. The most similar studies in terms of inputs and outputs are \cite{Reichmann2004}, \cite{Miidla2009} and \cite{Shahwan2013}. Our paper is, however, unique in the sample size -- we analyze \num{4660} municipal libraries in total. For comparison, the average sample size is 73 in the 16 studies we review in Table \ref{tab:lit}. Such a large data sample allows us to thoroughly investigate the impact of the operating environment on performance of libraries. We consider three possible environmental variables -- population of municipality, population density and distance to municipality with extended powers\footnote{The Czech Republic is divided into 8 cohesion regions (NUTS 2 -- region soudržnosti), 14 regions (NUTS 3 -- kraj), 77 districts (LAU 1 -- okres), 206 municipalities with extended powers (obec s rozšířenou působností), 393 municipalities with authorized municipal office (obec s pověřeným obecním úřadem) and \num{6258} municipalities (LAU 2 -- obec) as of April 1, 2019.}. Using regression analysis, we find that the efficiency score is significantly increasing with population. Extremely small villages are the exception as they tend to have higher efficiency score than villages with slightly higher population due to their very low and often zero inputs. We also find that for smaller villages the efficiency score is decreasing with distance to municipality with extended powers. Population density is insignificant in our analysis. Motivated by these results, we split the sample of libraries into 11 categories using decision tree analysis. We perform DEA separately for each category filtering out the influence of heterogeneous operating environment. This also decreases the discriminatory power of DEA which is very high in the preliminary analysis due to large sample size. The effect of distance is removed but the effect of population is not completely eliminated although it is reduced. This means that the distance can be safely treated as environmental variable while the population requires a more cautious approach as it is partially environmnetal and partially explanatory variable. Our proposed separation approach is quite suitable for this situation in contrast to the all-in-one, two-stage and multi-stage models that would take population as strictly environmental variable (see e.g.\ \citealp{Yang2009, DeWitte2010}). We also perform DEA for expert-defined categories and find that the proposed separation approach is robust to specification of subsamples to a certain degree. Our study contributes to the field of two-stage efficiency analysis -- one of the four active research fronts in DEA according to \cite{Liu2016a}.

The rest of the paper is structured as follows. In Section \ref{sec:lit}, we review the literature dealing with DEA and efficiency of libraries. In Section \ref{sec:meth}, we describe the Chebyshev distance DEA model used in the first stage and the regression model with decision tree model for analyzing efficiency scores used in the second stage. In Section \ref{sec:emp}, we compute efficiency scores of Czech public libraries in the year 2017 and investigate the impact of the operating environment. We conclude the paper in Section \ref{sec:con}.

\section{Literature Review}
\label{sec:lit}

\subsection{Data Envelopment Analysis}
\label{sec:litDea}

Data envelopment analysis (DEA) is a non-parametric method for the estimation of the production frontier (or, more precisely, the best-practice frontier) introduced by \cite{Charnes1978}. It measures technical efficiency of a decision making unit (DMU) relatively to the other units in the sample. The units that form the frontier are classified as efficient while the units not on the frontier are considered as inefficient. Inefficient units are further assigned efficiency score measuring their shortcomings. The efficiency classification as well as the efficiency score is determined based on how efficiently can a unit transform a set of inputs to a set of outputs. The original model of \cite{Charnes1978} denoted as the CCR model utilizes the constant returns to scale (CRS), i.e.\ it is assumed that an increase in inputs results in a proportionate increase in outputs. Variable returns to scale (VRS) relax this assumption and are utilized in the model of \cite{Banker1984} denoted as the BCC model. Many more models are proposed in the literature addressing various issues in DEA. A particulary convenient and elegant model is the Chebyshev distance model of \cite{Hladik2019}. It is based on the robust optimization viewpoint and has many attractive properties such as the super-efficiency, i.e.\ ability to assign scores to efficient units, and natural normalization, i.e.\ comparability of efficiency scores across different analyzes. For a survey of the DEA theory, see \cite{Cook2009}.

DEA is a very popular benchmarking tool in operations research and has a wide range of applications including but not limited to banking \citep{Fukuyama2017}, business \citep{Shabani2019}, agriculture \citep{Atici2015}, transportation \citep{Wu2016a}, health care \citep{Ozcan2017}, education \citep{Jablonsky2016}, research  \citep{Holy2018e} and sport \citep{Jablonsky2018}. For a survey of DEA applications, see \cite{Liu2013}.

Procedures for the practical use of DEA with its pitfalls are presented in \cite{Golany1989}, \cite{Boussofiane1991}, \cite{Dyson2001} and \cite{Cook2014}. One particular issue many studies face is heterogeneous operating environment. For DEA to make sense, however, the operating environment should be homogeneous. There exist several approaches for dealing with heterogeneous operating environment in the literature. For a review of such methods, see \cite{Yang2009} and \cite{DeWitte2010}. We briefly describe the four most commonly used methods for DEA. The separation approach splits the heterogeneous data sample into several homogeneous subsamples according to one or more environmental variables and performs DEA separately for each subsample. The advantage of this approach is its simplicity and straightforward interpretability. However, it significantly reduces the sample size making it unusable in many studies. The all-in-one model directly includes environmental variables in DEA as inputs or outputs. The two-stage model adjusts the efficiency scores based on the dependence between preliminary efficiency scores and environmental variables using regression analysis. The multi-stage model regress input slacks on environmental variables, adjusts inputs and finally performs DEA with adjusted inputs. The latter three models are more sophisticated and do not reduce sample size but are more cumbersome to interpret.

Whether theoretical, applicational or practical, the literature dealing with DEA is very extensive and still growing. \cite{Emrouznejad2018} report a listing of scientific articles related to DEA from the seminal paper of \cite{Charnes1978} to 2016. \cite{Liu2016a} identify the research activities (or the research fronts) in DEA from 2000 to 2014.

\subsection{Efficiency of Libraries}
\label{sec:litLib}

One of the possible uses of DEA is assesing the efficiency of public or university libraries in a given area at a given time. We review 16 papers dealing with efficiency of libraries. The overview of papers is presented in Table \ref{tab:lit}. Most studies utilize the classical CCR or BCC DEA models although some studies adopt free disposal hull (FDH) approach. \cite{Simon2011} and \cite{Guccio2018} consider intermediate outputs and adopt network DEA with two steps. \cite{Witte2011} focus only on the first step that produces intermediate outputs. We compare all 16 studies based on the sample size, selection of the inputs and outputs and treatment of the operating environment.

In the literature, public libraries are analyzed in the order of tens or hundreds at most. The largest sample size among the reviewed studies is 290 \citep{Witte2011} followed by 184 \citep{Vitaliano1998}, 118 \citep{Reichmann2004}, 99 \citep{Hammond2002} and 92 \citep{Vrabkova2019}. The rest of the studies have quite small sample size ranging from 11 to 68.

The reviewed studies utilize up to 5 inputs and up to 4 outputs. The most common inputs are the number of employees or personnel expenditures (87.50 percent of studies), book or other collections (62.50 percent of studies), variables related to expenditures (56.25 percent of studies) and the area of library (37.50 percent of studies). The most common outputs are the circulation or the number of loans (93.33 percent of studies), the number of visits (40.00 percent of studies), the number of consultations (40.00 percent of studies) and the number of registrations (33.33 percent of studies). The number of additions to collection, the opening hours and the number of serial subscriptions appear less often in the literature and in some studies are considered as inputs while in others as outputs or intermediate outputs.

Some of the studies consider operating environment to a certain degree. \cite{Sharma1999}, \cite{Reichmann2004}, \cite{Chen2005}, \cite{Miidla2009}, \cite{Reichmann2010}, \cite{Stroobants2014} and \cite{Vrabkova2019} analyze behavior of libraries in several predefined groups and compare their efficiency scores. \cite{Srakar2017} follow a similar approach but cluster libraries according to their efficiency and size with additional spatial constraints. \cite{Vitaliano1998} uses the tobit regression to model efficiencies and find that they are positively dependent on population, negatively on wages of the directors and positively on town or village associations. \cite{Hammond2002} includes population density and accessibility measures in the DEA model as non-discretionary inputs. \cite{Witte2011} employ the conditional efficiency model and find that the efficiency increases with left-wing ideological stance of the local government, wealth of the population, population density and local funding.

\begin{center}
\begin{longtable}{rl}
\caption{Overview of relevant studies.}
\label{tab:lit}
\endfirsthead
\toprule
Paper:   & \cite{Chen1997} \\
Sample:  & 23 University Libraries in Taipei, Taiwan \\
Inputs:  & Operating Expenditures, Employees, Area \\
Outputs: & Visits, Circulation, Inter-Library Circulation, Consultations \\
\midrule
Paper:   & \cite{Vitaliano1998} \\
Sample:  & 184 Public Libraries in New York, United States \\
Inputs:  & Collection, Collection Additions, Serial Subscriptions, Opening Hours \\
Outputs: & Circulation, Consultations \\
\midrule
Paper:   & \cite{Sharma1999} \\
Sample:  & 47 Public Libraries in Hawaii, United States. \\
Inputs:  & Operating Expenditures, Employees, Collection, Days Open \\
Outputs: & Visits, Circulation, Consultations \\
\midrule
Paper:   & \cite{Hammond2002} \\
Sample:  & 99 Public Libraries in the United Kingdom \\
Inputs:  & Collection, Collection Additions, Serial Subscriptions, Opening Hours \\
Outputs: & Circulation, Consultations, Requests \\
\midrule
Paper:   & \cite{Reichmann2004} \\
Sample:  & 118 University Libraries in English-Speaking and German-Speaking Countries \\
Inputs:  & Employees, Collection \\
Outputs: & Circulation, Opening Hours, Collection Additions, Serial Subscriptions \\
\midrule
Paper:   & \cite{Chen2005} \\
Sample:  & 23 Public Libraries in Tokyo, Japan \\
Inputs:  & Employees, Collection, Area, Population \\
Outputs: & Registrations, Circulation \\
\midrule
Paper:   & \cite{Miidla2009} \\
Sample:  & 20 Central Public Libraries in Estonia \\
Inputs:  & Operating Expenditures, Personnel Expenditures, Collection, Area \\
Outputs: & Registrations, Circulation \\
\midrule
Paper:   & \cite{Reichmann2010} \\
Sample:  & 68 University Libraries in North America, Austria and Germany \\
Inputs:  & Employees, Collection \\
Outputs: & Circulation, Collection Additions, Serial Subscriptions \\
\midrule
Paper:   & \cite{Witte2011} \\
Sample:  & 290 Municipal Public Libraries in Flanders, Belgium \\
Inputs:  & Operating Expenditures, Personnel Expenditures, Infrastructure Expenditures \\
Inter.:  & Youth Book Collection, Book Collection, Media Collection, Opening Hours \\
\midrule
Paper:   & \cite{Simon2011} \\
Sample:  & 34 University Libraries in Spain \\
Inputs:  & Operating Expenditures, Employees, Area \\
Inter.:  & Collection, Serial Subscriptions, Opening Hours, Seats \\
Outputs: & Circulation, Inter-Library Circulation, Downloads \\
\midrule
\newpage
\midrule
Paper:   & \cite{DeCarvalho2012} \\
Sample:  & 37 University Libraries in Rio de Janeiro, Brazil \\
Inputs:  & Employees, Collection, Area \\
Outputs: & Registrations, Visits, Circulation, Consultations \\
\midrule
Paper:   & \cite{Shahwan2013} \\
Sample:  & 11 Academic Libraries in the Arab States of the Gulf \\
Inputs:  & Total Expenditures, Employees, Collection \\
Outputs: & Registrations, Circulation, Collection Additions \\
\midrule
Paper:   & \cite{Stroobants2014} \\
Sample:  & 13 Local Public Libraries in Flanders, Belgium \\
Inputs:  & Total Expenditures / Operating Expenditures, Employees \\
Outputs: & Circulation / Circulation, Opening Hours \\
\midrule
Paper:   & \cite{Srakar2017} \\
Sample:  & 58 Public General Libraries in Slovenia \\
Inputs:  & Total Expenditures, Employees, Area, Ratio of Service Points to Potential Users \\
Outputs: & Registrations, Visits / Circulation / Equipment / Events, Events Attendance \\
\midrule
Paper:   & \cite{Guccio2018} \\
Sample:  & 44 Public State Libraries in Italy \\
Inputs:  & Non-Personnel Expenditures, Employees, Shelf Size, Seats. \\
Inter.:  & Book, Manuscript, Periodical and Other Collections, Assets Value. \\
Outputs: & Visits, Circulation, Inter-Library Circulation, Consultations \\
\midrule
Paper:   & \cite{Vrabkova2019} \\
Sample:  & 92 Public Libraries in the Czech Republic and Slovakia \\
Inputs:  & Employees, Collection, Collection Additions, Events, Opening Hours \\
Outputs: & Visits \\
\bottomrule
\end{longtable}
\end{center}

\section{Methodology}
\label{sec:meth}

\subsection{Chebyshev Distance Data Envelopment Analysis}
\label{sec:methDea}

To obtain technical efficiencies, we utilize the Chebyshev distance DEA with variable returns to scale (VRS) proposed by \cite{Hladik2019}. Let $X = (x_{i,j})_{i=1,k=1}^{n,r}$ be the non-negative matrix of inputs and $Y = (y_{i,k})_{i=1,k=1}^{n,s}$ be the non-negative matrix of outputs. We denote $x_{i} = (x_{i,1}, \ldots, x_{i,r})'$ and $y_{i} = (y_{i,1}, \ldots, y_{i,s})'$ the vectors corresponding to the $i$-th row. We also denote $X_{-i}$ and $Y_{-i}$ the matrices with $i$-th row missing, i.e. the inputs and outputs of every DMU but $i$.

As in the classical DEA models, the problem of measuring efficiency of a DMU is formulated as finding the optimal weights of input and output variables with respect to the other DMUs. Note that each DMU has its own optimization problem. The idea of the Chebyshev distance DEA is to rank DMUs based on robustness of efficiency or inefficiency classification to variations of input and output data using the Chebyshev distance. Specifically, the resulting efficiency score for $i$-th DMU is equal to $r_{i} = 1 + 2 \delta_{i}^*$, where $\delta_{i}^*$ is the optimal solution to the optimization problem
\begin{equation}
\label{eq:hladikNon}
\begin{aligned}
\max_{\substack{\delta_{i}, \nu_{i}, \mu_{i}, \varphi_{i}}} && \multispan2{$\delta_{i}$} \\
\text{such that} && (1 - \delta_{i}) y_{i}' \mu_{i} - \varphi_{i} & \geq 1, \\
&& (1 + \delta_{i}) x_{i}' \nu_{i} & \leq 1, \\
&& (1 + \delta_{i}) Y_{-i} \mu_{i} - (1 - \delta_{i}) X_{-i} \nu_{i} - 1 \varphi_{i} & \leq 0, \\
&& \mu_{i} & \geq 0, \\
&& \nu_{i} & \geq 0, \\
\end{aligned}
\end{equation}
where $\nu_{i} = (\nu_{i,1}, \ldots, \nu_{i,r})'$ are the weights of inputs, $\mu_{i} = (\nu_{i,1}, \ldots, \nu_{i,s})'$ are the weights of outputs and $\varphi_{i}$ is the auxiliary variable used for ensuring VRS. The above formulation is a non-linear optimization problem which \cite{Hladik2019} further propose to linearize. Let us reparametrize the weights and the VRS variable as
\begin{equation}
\tilde{\nu}_{i} = \frac{\nu_{i}}{1 + \delta}, \qquad
\tilde{\mu}_{i} = \frac{\mu_{i}}{1 - \delta}, \qquad
\tilde{\varphi}_{i} = \frac{\varphi_{i}}{1 - \delta^2}.
\end{equation}
The linear approximation of \eqref{eq:hladikNon} is then given by
\begin{equation}
\label{eq:hladikLin}
\begin{aligned}
\max_{\substack{\delta_{i}, \tilde{\nu}_{i}, \tilde{\mu}_{i}, \tilde{\varphi}_{i}}} && \multispan2{$\delta_{i}$} \\
\text{such that} && y_{i}' \tilde{\mu}_{i}  - \tilde{\varphi}_{i} & \geq 1 + 2 \delta_{i}, \\
&& x_{i}' \tilde{\nu}_{i} & \leq 1 - 2 \delta_{i}, \\
&& Y_{-i} \tilde{\mu}_{i} - X_{-i} \tilde{\nu}_{i} - 1 \tilde{\varphi}_{i} & \leq 0, \\
&& \tilde{\mu}_{i} & \geq 0, \\
&& \tilde{\nu}_{i} & \geq 0. \\
\end{aligned}
\end{equation}
\cite{Hladik2019} shows in several examples that the linear approximation \eqref{eq:hladikLin} is quite precise and can be effectively utilized in practice.

The efficiency scores $r_{i}$, $i=1,\ldots,n$ lie in interval $[0, 2]$ whether given by the original non-linear optimization problem \eqref{eq:hladikNon} or its linear approximation \eqref{eq:hladikLin}. Values $r_{i} \in [0,1)$ indicate inefficient DMUs while values $r_{i} \in [1,2]$ indicate efficient DMUs. The Chebyshev distance DEA further possesses the following properties:
\begin{itemize}
\item \textbf{Robust Interpretation}: The efficiency scores of the Chebyshev distance DEA indicate how DMUs are sensitive to changes in their inputs and outputs. Specifically, the efficiency scores for inefficient DMUs are the smallest possible variations of all inputs and outputs causing efficiency in terms of the Chebyshev distance while the efficiency scores for efficient DMUs are the largest possible variations of all inputs and outputs preserving efficiency.
\item \textbf{Super-Efficiency}: As noted above, the Chebyshev distance DEA ranks inefficient as well as efficient DMUs. In contrast, the basic formulation of the classical DEA allows only for ranking of inefficient DMUs.
\item \textbf{Normalization}: The efficiency scores of the Chebyshev distance DEA are naturally normalized due to their robust interpretation. Therefore, the efficiency scores can be compared across different analyzes.
\item \textbf{Non-Negativeness}: Unlike classical DEA, the Chebyshev distance DEA allows for zero inputs and zero outputs as well.
\item \textbf{Units Invariance}: Similarly to the classical DEA, the inputs and outputs can be arbitrarily scaled without affecting the efficiency scores of the Chebyshev distance DEA model. Therefore, it does not matter in which units are the inputs and outputs measured. 
\item \textbf{Ranking Order}: The classification to efficient and inefficient DMUs as well as the order of inefficient DMUs according to their efficiency score is exactly the same in the Chebyshev distance DEA model as in the classical CCR model (or the BBC model when assuming VRS). The values of the efficiency scores, however, differ.
\end{itemize}

\subsection{Analysis of Efficiency Scores in the Second Stage}
\label{sec:methReg}

We utilize the linear regression for modeling efficiency scores in a similar fashion as \cite{Holy2018e}. Let $m$ be the number of regressors and $Z = (z_{i,l})_{i=1,l=1}^{n,m}$ the design matrix with the values of the regressors. We further denote $z_i = (z_{i,1},\ldots,z_{i,m})'$ the vector corresponding to the $i$-th row of $Z$. As efficiency scores of the Chebyshev distance DEA $r_i$ are bounded from bellow by 0 and from above by 2, we resort to the regression model with the logistic function
\begin{equation}
\label{eq:regLogit}
r_i = \frac{2}{1 + e^{- z'_i \beta - \varepsilon_i}}, \qquad \varepsilon_i \stackrel{iid}{\sim} \mathrm{N} \left( 0, \sigma^2 \right), \qquad i = 1, \ldots, n,
\end{equation}
where $\beta = (\beta_1, \ldots, \beta_m)'$ and $\sigma^2$ are the unknown parameters. Next, we use the transformation $\tilde{r}_i = \ln \left( r_i / (2 - r_i) \right)$ and arrive at the linear regression model
\begin{equation}
\label{eq:regLin}
\tilde{r}_i = z'_i \beta + \varepsilon_i, \qquad \varepsilon_i \stackrel{iid}{\sim} \mathrm{N} \left( 0, \sigma^2 \right), \qquad i = 1, \ldots, n.
\end{equation}

Note that we assume that $\varepsilon_i$ are independent. This is clearly not the case as there is inherent dependency between the efficiency scores obtained by DEA. Serial correlation affects mainly the inference while the estimate of coefficients $\beta$ remains unbiased and consistent. As studied by \cite{Simar2007}, the dependency structure is complex and unknown but disappears asymptotically. Our data sample is quite large and we therefore resort to the independence simplification as most studies.

We further analyze efficiency scores using the decision tree approach. To build the decision tree, we adopt the RPART routine of \cite{Therneau2019}. Again, we analyze dependency of the efficiency scores $r_i$ on the regressors $z_i$, $i=1,\ldots,n$.

\section{Empirical Study}
\label{sec:emp}

\subsection{Data Sample}
\label{sec:empData}

We analyze efficiency of public libraries established by Czech municipalities during the year 2017. In total, there are \num{5339} public libraries in 2017. Of these, \num{4790} are established by municipalities excluding Prague and \num{38} by municipal and administrative districts of Prague. The remaining libraries include the National Library of the Czech Republic, the Moravian Library in Brno, the 13 regional libraries, libraries established by districts, etc. We focus only on the municipal libraries outside the capital. In our data, 2.71 percent of libraries have some observations missing. We remove these libraries from the analysis. Our data sample therefore consists of \num{4660} municipal libraries with no missing data. We have data available for the years 2016 and 2017. The two year history allows us to utilize aggregated values and first differences in the analysis.

For in-depth statistics about public libraries in the Czech Republic, we refer to the National Information and Consulting Centre for Culture (NIPOS).

\subsection{Variable Selection}
\label{sec:empVar}

\begin{figure}
\begin{center}
\includegraphics[width=0.9\textwidth]{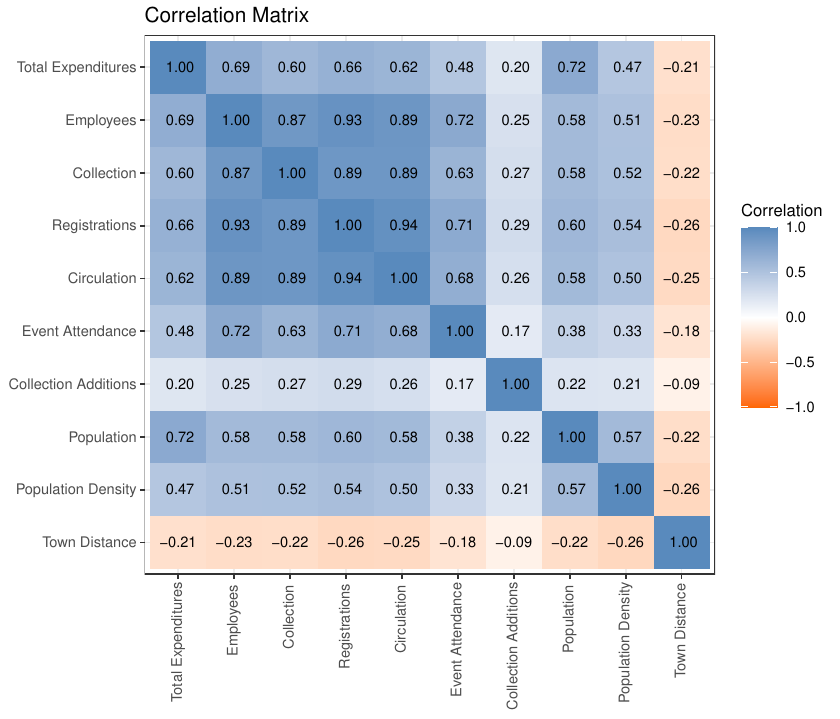}
\caption{Correlation matrix of the input, output and environmental variables.}
\label{fig:cor}
\end{center}
\end{figure}

In our study, we utilize 10 variables in total. Descriptive statistics of the variables are reported in Table \ref{tab:desc}. The correlation matrix is illustrated in Figure \ref{fig:cor}. All variables except the town distance are strongly positively correlated while the town distance is moderately negatively correlated with the others. For the efficiency analysis, we consider the following $r=3$ input variables:
\begin{itemize}
\item \textbf{Total Expenditures}: The total expenditures in CZK by the municipality to library activities (class 3314 in the sectoral classification of budget structure) in 2016 and 2017. We aggregate the expenditures to two years to capture long-term investments and smooth out annual budget changes. In 2016 and 2017, 6.78 percent of libraries did not receive any funding specifically for the operation of libraries as very small libraries can be part of the municipal office and share its budget. The data source is information portal MONITOR of the Ministry of Finance of the Czech Republic.
\item \textbf{Employees}: The number of full-time equivalents of library employees in 2017. Note that 64.08 percent of libraries have no own employees as very small libraries are run either by employees of the municipal office or volunteers. The data source is NIPOS.
\item \textbf{Collection}: The total number of book units owned by the library in 2016. This variable represents the capital of the library. We use the value from the previous year as we consider the increase in book collection in the current year to be output variable reflecting the performance of the library management. A small number of libraries report no book collection. Nevertheless, we keep them in the dataset as the used methodology can deal with this situation. The data source is NIPOS.
\end{itemize}
We denote the input variables respectively as $x_{i, 1}$, $x_{i, 2}$ and $x_{i, 3}$, $i = 1, \ldots, n$. Further inputs such as the area of the library, the equipment, more detailed expenditures or more detailed collections could also be utilized. Unavailability of these variables is not a fundamental problem as the three major input groups -- operating costs, personnel and capital assets -- are represented in our study. However, we measure capital assets only by the number of books and, due to our data limitations, we are forced to omit media collections including e-books and audio-visual materials which play an important role in modern libraries.

We consider the following $s=4$ output variables:
\begin{itemize}
\item \textbf{Registrations}: The total number of users registered in the library in 2017. This variable captures the size of the reader base. The data source is NIPOS.
\item \textbf{Circulation}: The total number of book loans in 2017. This variable captures the main activity of libraries -- book lending. The data source is NIPOS.
\item \textbf{Events Attendance}: The total number of visitors of events organized by the library in 2017. This variable captures the cultural role of libraries. Many libraries do not organize any events while others offer regular cultural program. The data source is NIPOS.
\item \textbf{Collection Additions}: The positive part of difference between the book collection in 2017 and 2016. This variable captures the increase of the capital of libraries. According to Table \ref{tab:desc}, the book collection of 50.56 percent libraries remains the same as in 2016 or in some cases even decreases. The data source is NIPOS.
\end{itemize}
We denote the output variables respectively as $y_{i, 1}$, $y_{i, 2}$, $y_{i, 3}$ and $y_{i, 4}$, $i = 1, \ldots, n$. Further outputs such as the number of visits, the number of consultations, the opening hours, the inter-library circulation or various measures of the internet activity could also be utilized. Unfortunately, we do not have these variables available in our data. Notably, we are missing outputs related to the presence usage and information services. Our variables therefore capture only the more traditional role of libraries.

Finally, we consider the folowing 3 variables potentially describing the environment in which libraries operate:
\begin{itemize}
\item \textbf{Population}: The number of inhabitants of the municipality as of January 1, 2018. The data source is the Czech Statistical Office (CSO). We denote this variable as $p_{i}$, $i = 1,\ldots,n$.
\item \textbf{Population Density}: The number of inhabitants of the municipality per square kilometre as of January 1, 2018. The data source is CSO. We denote this variable as $d_{i}$, $i = 1,\ldots,n$.
\item \textbf{Town Distance}: The travel time by car in minutes to the municipality with extended powers\footnote{We have also considered different specifications of distance and reference town. Instead of the travel time, we have tried the air distance and road distance. Instead of the municipality with extended powers (obec s rozšířenou působností), we have tried the district capital (LAU 1 -- okresní město), regional capital (NUTS 3 -- krajské město), town with population higher than \num{10000} and city with general significance. All combinations of distances and reference towns have lead to weaker results.}. The data source is Czech web mapping service Mapy.cz. We denote this variable as $t_{i}$, $i = 1,\ldots,n$.
\end{itemize}

\begin{table}
\begin{center}
\caption{Descriptive statistics of the input, output and environmental variables.}
\label{tab:desc}
\begin{tabular}{llrrrrr}
\toprule
Not. & Variable & Mean & Std. Dev. & Maximum & Zeros \\
\midrule
$x_{i,1}$ & Total Expenditures   & \num{374175.72} & \num{1597571.60} & \num{42799631.00} &  \num{6.78}\% \\
$x_{i,2}$ & Employees            &      \num{0.34} &       \num{1.21} &       \num{20.66} & \num{64.08}\% \\
$x_{i,3}$ & Collection           &   \num{5477.19} &   \num{10836.46} &   \num{166139.00} &  \num{2.94}\% \\
$y_{i,1}$ & Registrations        &    \num{126.69} &     \num{334.49} &     \num{4673.00} &  \num{0.09}\% \\
$y_{i,2}$ & Circulation          &   \num{4750.39} &   \num{15779.77} &   \num{236983.00} &  \num{0.11}\% \\
$y_{i,3}$ & Event Attendance     &    \num{357.87} &    \num{1593.40} &    \num{43972.00} & \num{52.42}\% \\
$y_{i,4}$ & Collection Additions &     \num{70.54} &     \num{313.89} &    \num{15625.00} & \num{50.56}\% \\
$p_{i}$   & Population           &   \num{1327.29} &    \num{3881.99} &   \num{103979.00} &  \num{0.00}\% \\
$d_{i}$   & Population Density   &     \num{98.62} &     \num{129.45} &     \num{2115.83} &  \num{0.00}\% \\ 
$t_{i}$   & Town Distance        &     \num{14.68} &       \num{7.40} &       \num{58.12} &  \num{2.96}\% \\
\bottomrule
\end{tabular}
\end{center}
\end{table}

\begin{figure}
\begin{center}
\includegraphics[width=0.9\textwidth]{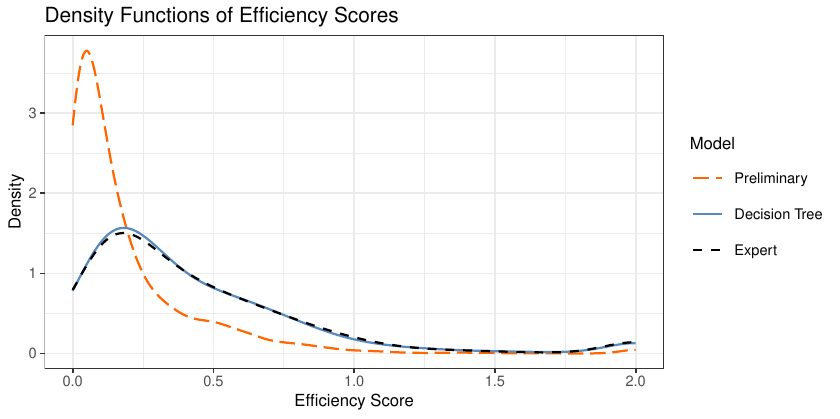}
\caption{Kernel density functions of the efficiency scores.}
\label{fig:density}
\end{center}
\end{figure}

\subsection{Preliminary Efficiency Analysis}
\label{sec:empPre}

First, we apply the presented Chebyshev distance DEA with selected inputs and outputs to the full dataset of \num{4660} libraries. We denote this as the preliminary efficiency analysis. Note that we consider VRS as there are huge differences in sizes of libraries and we do not assume proportional changes in inputs and outputs. Returns to scale can then be either increasing, decreasing or even constant. The estimated density function of preliminary efficiency scores is illustrated in Figure \ref{fig:density}. For the estimation of the density, we utilize the Gaussian kernel. As expected for such large dataset, most libraries are inefficient with very low efficiency score. Specifically, 98.45 percent of all units are inefficient with mean score 0.1916 and median score 0.0999. Note that the mean efficiency score should not be interpreted in terms of efficiency/inefficiency. Instead, it indicates the overall performance level as higher efficiency score always means better performing unit.

In the next steps, we improve this preliminary approach and focus on two issues -- the operational environment and the discriminatory power. We investigate whether our sample of units is homogeneous (i.e.\ all libraries operate within the same environment) or heterogeneous (i.e.\ libraries operate under different conditions). Based on our findings, we divide the full sample into several smaller categories according to the environmental influences. This not only ensures homogeneity but also reduces the overly strict discriminatory power.

\subsection{Dependence on Explanatory Variables}
\label{sec:empDep}

We study the influence of the population $p_i$, the population density $d_i$ and the town distance $t_i$ on the transformed preliminary efficiency score $\tilde{r}_i$ of the unit $i=1,\ldots,n$ using the linear regression. We arrive at the model formulation\footnote{Before arriving at this final model, we have tried several specifications of the regression model including all variables $p_i$, $d_i$ and $t_i$ with logarithmic and power transformations as well as various interactions.}
\begin{equation}
\tilde{r}_{i} = \beta_0 + \beta_1 \ln(p_i) + \beta_2 \frac{1}{\ln(p_i)} + \beta_3 \frac{t_i}{p_i} + \varepsilon_i, \qquad \varepsilon_i \stackrel{iid}{\sim} \mathrm{N} (0, \sigma^2), \qquad i = 1,\ldots,n,
\end{equation}
where $\beta_0$, $\beta_1$, $\beta_2$, $\beta_3$ and $\sigma^2$ are the parameters. Results of the regression model are reported in Table \ref{tab:reg}. For the preliminary efficiency scores, all regressors are statistically significant at any reasonable confidence level. The model explains 22.90 percent of variance in the dependent variable.

The above regression model has the following interpretation. The efficiency score increases with population as the coefficient $\beta_1$ is positive. For very small population, however, the efficiency score also increases as the coefficient $\beta_2$ is also positive. Finally, the efficiency score increases with decreasing town distance as the coefficient $\beta_3$ is negative. This relation is more distinctive for smaller population as the town distance $t_i$ is divided by the population $p_i$. We do not include the population density $d_i$ in the final model as it is not significant in any transformation.

The regression model describes the relationship between the efficiency score and possible environmental factors. However, it does not tell us whether the population and town distance cause change in the efficiency and can be considered as environmental factors.

\begin{table}
\begin{center}
\caption{Summary of regression models.}
\label{tab:reg}
\begin{tabular}{lllrrrr}
\toprule
Model & Coeff. & Regressor & Estimate & Std. Error & t-Statistic & p-Value \\
\midrule
\multirow{2}{*}{Preliminary}         & $\beta_0$ & Intercept    & -24.0894 & 1.4866 & -16.2049 & 0.0000 \\
                                     & $\beta_1$ & $\ln(p_i)$   &   1.9496 & 0.1082 &  18.0230 & 0.0000 \\
\multirow{2}{*}{$R^2 = 0.2290$}      & $\beta_2$ & $1/\ln(p_i)$ &  54.3415 & 5.1511 &  10.5495 & 0.0000 \\
                                     & $\beta_3$ & $t_i/p_i$    &  -2.7975 & 0.7090 &  -3.9456 & 0.0001 \\
\midrule
\multirow{2}{*}{Decision Tree}       & $\beta_0$ & Intercept    & -15.6972 & 2.2017 &  -7.1295 & 0.0000 \\
                                     & $\beta_1$ & $\ln(p_i)$   &   1.3447 & 0.1602 &   8.3933 & 0.0000 \\
\multirow{2}{*}{$R^2 = 0.0600$}      & $\beta_2$ & $1/\ln(p_i)$ &  35.5161 & 7.6292 &   4.6553 & 0.0000 \\
                                     & $\beta_3$ & $t_i/p_i$    &  -0.2859 & 1.0501 &  -0.2723 & 0.7854 \\            
\midrule
\multirow{2}{*}{Expert}              & $\beta_0$ & Intercept    & -21.8288 & 2.2922 &  -9.5233 & 0.0000 \\
                                     & $\beta_1$ & $\ln(p_i)$   &   1.8568 & 0.1668 &  11.1323 & 0.0000 \\
\multirow{2}{*}{$R^2 = 0.0834$}      & $\beta_2$ & $1/\ln(p_i)$ &  53.6474 & 7.9426 &   6.7544 & 0.0000 \\
                                     & $\beta_3$ & $t_i/p_i$    &  -1.0758 & 1.0932 &  -0.9841 & 0.3251 \\
\bottomrule
\end{tabular}
\end{center}
\end{table}

\subsection{Efficiency Analysis with Decision Tree Categories}
\label{sec:empTree}

The regression model indicates dependency of the efficiency score on the population and town distance. We further support this claim by decision tree analysis. Other motivation for the use of the decision tree is separation of the data sample to several subsamples. As our goal is to use subsamples for separate efficiency analysis, we want them to have roughly the same number of units. Unfortunately, this is not guaranteed by the decision tree and we must therefore control the building of the tree by restricting the minimum number of units in a category. We find that in our case, the minimum of \num{138} units leads to the most interpretable results. Another tuning parameter is the number of categories or the depth of the tree. We find that 11 categories with depth 7 is an adequate choice.

\begin{figure}
\begin{center}
\includegraphics[width=0.9\textwidth]{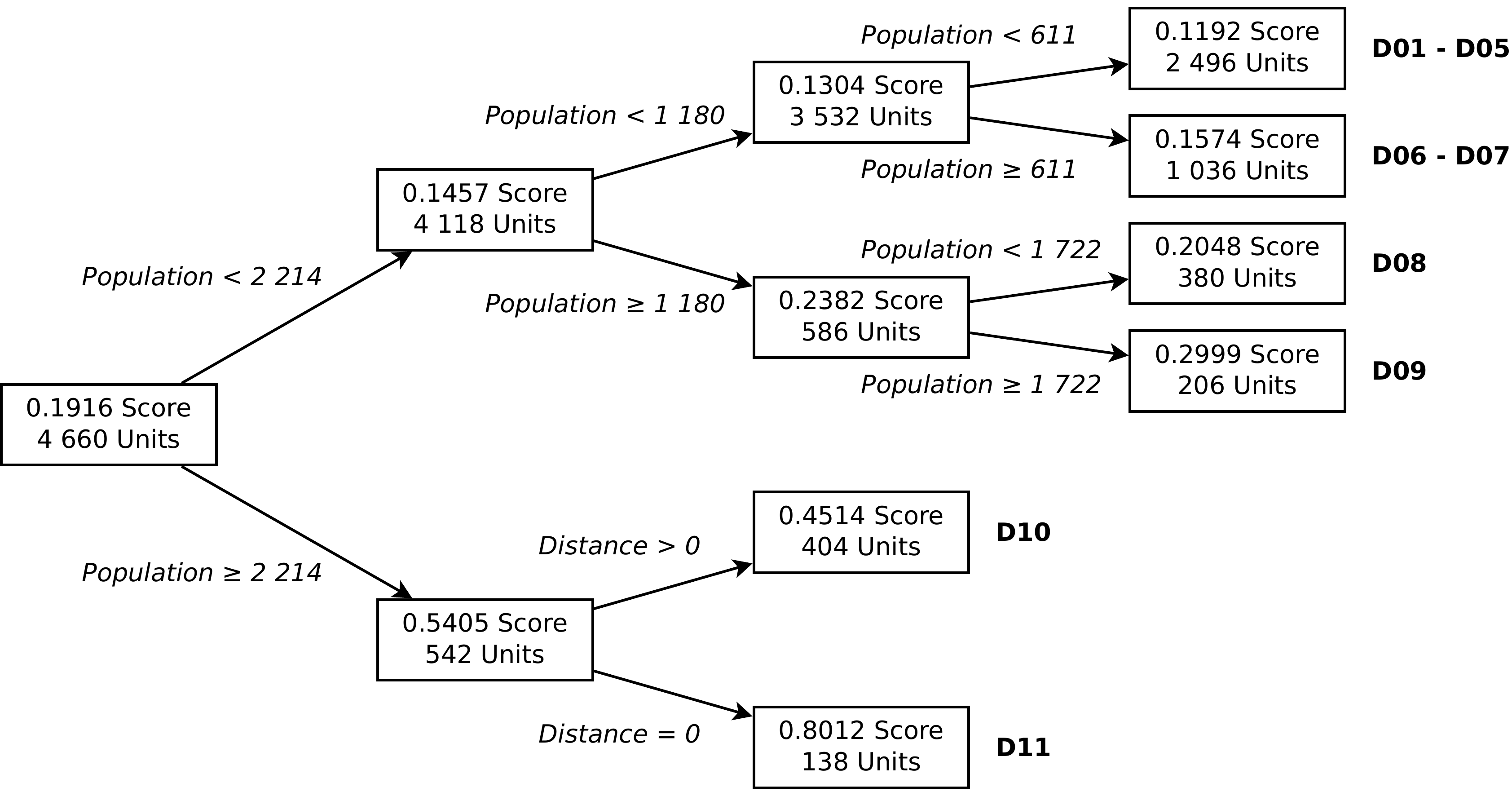}
\caption{Decision tree of depth 3 with mean efficiency scores and the numbers of units.}
\label{fig:tree}
\end{center}
\end{figure}

The categories of libraries given by the decision tree together with mean values of preliminary efficiency scores are reported in Table \ref{tab:effTree}. We denote the categories as D01--D11. The decision tree divides the units into small with population lower than \num{611} (categories D01--D05), medium with population between \num{611} and \num{2214} (categories D06--D09) and large with population higher than \num{2214} (categories D10 and D11). Small units are further divided according to the town distance, medium according to the population and large to municipalities with extended powers (category D11) and other towns (category D10). As in the regression model, the town distance is more important for the smaller units. However, the mean efficiency scores suggest that the relation might be more complex -- likely due to dependence between population and town distance. Decision tree also finds that it is significant whether the town distance is zero (and the unit is therefore the reference town) or positive as it puts all municipalities with extended powers into the category D11. The building of the decision tree is illustrated in Figure \ref{fig:tree}.

\begin{figure}
\begin{center}
\includegraphics[width=0.9\textwidth]{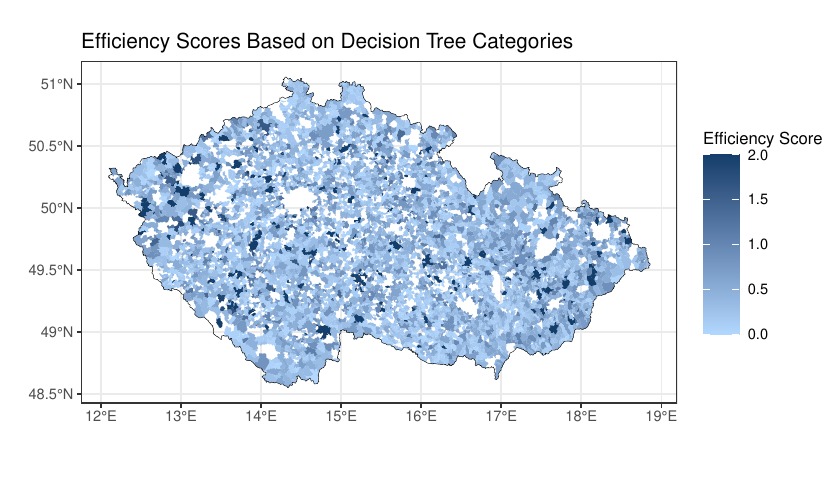}
\caption{Efficiency scores based on decision tree categories for each municipality.}
\label{fig:map}
\end{center}
\end{figure}

Next, we calculate efficiency scores separately for each category given by the decision tree. The mean efficiency scores are reported in Table \ref{tab:effTree}. The efficiency scores for each municipality are illustrated in Figure \ref{fig:map}. The discriminatory power of this efficiency analysis is more reasonable as 92.30 percent of all units are inefficient with mean score 0.4371 and median score 0.3070. The shape of the score density function is relatively mild as illustrated in Figure \ref{fig:density}. Note that the preliminary scores have different interpretation than the decision tree scores as they use different samples. For example the fact that the mean decision tree score of D05 is higher than the mean score of D04 does not imply that D05 is more efficient. On the contrary, preliminary scores show that D04 is on average more efficient. Only with the removal of D04 units and others from the efficiency analysis of D05, the D05 units become more efficient on average.

As for the preliminary scores, we use the regression model for the decision tree scores. Note that we are able to compare efficiency scores in different categories thanks to the normalization property of the Chebyshev distance DEA. Table \ref{tab:reg} shows that the town distance is no longer significant for the new scores. This suggests that the influence of the town distance is eliminated by the decision tree categories and the town distance is indeed an environmental factor. Our adjustment for the town distance in categories therefore leads to more fair comparison of libraries. The effect of the population, however, remains significant although it is a bit lower as the model explains only 6.00 percent of the efficiency scores variance. It is also evident from Table \ref{tab:effTree} that more units have higher efficiency score for categories with higher population. This suggests that the population have some partial environmental influence but we cannot attribute unilateral causal influence to it. Libraries in towns with larger population are simply far more efficient on average even if we treat smaller towns separately.

This is an important result advocating our separation approach. Unlike the all-in-one model, two-stage and multi-stage models, we do not consider exogenous variables to fully affect the operating environment. We use them to measure similarity between DMUs and then retain only similar DMUs in the data sample. Our approach therefore diminishes the environmental influence of dissimilar DMUs while keeping the unaltered influence of similar DMUs.

\begin{table}
\begin{center}
\caption{Mean efficiency scores within each decision tree category.}
\label{tab:effTree}
\begin{tabular}{lllrrrr}
\toprule
Cat. & Population & Distance & Units & Preliminary & Dec. Tree & Expert \\
\midrule
D01 & $[   \num{0},  \num{611})$ & $[  0.00,   8.63)$ & 373 & 0.1147 & 0.4893 & 0.3163 \\
D02 & $[   \num{0},  \num{611})$ & $[  8.63,  11.46)$ & 408 & 0.1413 & 0.3685 & 0.3630 \\
D03 & $[   \num{0},  \num{611})$ & $[ 11.46,  18.02)$ & 867 & 0.1077 & 0.2869 & 0.3047 \\
D04 & $[   \num{0},  \num{611})$ & $[ 18.02,  23.63)$ & 481 & 0.1451 & 0.2916 & 0.3924 \\
D05 & $[   \num{0},  \num{611})$ & $[ 23.63, \infty)$ & 367 & 0.0923 & 0.4609 & 0.2866 \\
D06 & $[ \num{611},  \num{667})$ & $[  0.00, \infty)$ & 165 & 0.1865 & 0.5353 & 0.4527 \\
D07 & $[ \num{667}, \num{1180})$ & $[  0.00, \infty)$ & 871 & 0.1519 & 0.3332 & 0.4265 \\
D08 & $[\num{1180}, \num{1722})$ & $[  0.00, \infty)$ & 380 & 0.2048 & 0.6471 & 0.6318 \\
D09 & $[\num{1722}, \num{2214})$ & $[  0.00, \infty)$ & 206 & 0.2999 & 0.7081 & 0.6497 \\
D10 & $[\num{2214},     \infty)$ & $(  0.00, \infty)$ & 404 & 0.4514 & 0.6130 & 0.7558 \\
D11 & $[\num{2214},     \infty)$ & $   0.00         $ & 138 & 0.8012 & 0.9256 & 0.9256 \\
\midrule
\multicolumn{3}{l}{All} & \num{4660} & 0.1916 & 0.4371 & 0.4458 \\
\bottomrule
\end{tabular}
\end{center}
\end{table}

\subsection{Efficiency Analysis with Expert Categories}
\label{sec:empExp}

The categorization by the decision tree is purely data-driven approach with its benefits and limitations. For example, it is a well known fact that decision trees are quite sensitive to changes in data and have tendency to overfit. We compare the categories given by the decision tree with categories selected by an expert. The expert categories can be useful in several ways. From the statistical point of view, their simpler rules can prevent sensitivity to data changes and offer more robust approach. From the applicability point of view, they can be used in variety of applications and time frames in contrast with our decision tree specifically designed for the efficiency analysis of public libraries in 2017. From the managerial point of view, it might be easier to convince management of the decision making units that expert categories with "nicer looking" rules are more fair. Nevertheless, the data-driven categories offer valuable insight and should serve as the benchmark.

Our expert categories with their rules are described in Table \ref{tab:effExp}. We keep the number of categories at 11 and denote them E01-E11. We divide units into 5 population levels and 2 distance levels forming 10 categories based on very simple rules with roughly the same size. We keep municipalities with extended powers in the separate category E11 identically to the decision tree category D11.

We follow the same procedure as for the efficiency analysis based on the decision tree. Efficiency scores within expert categories are reported in Table \ref{tab:effExp}. The discriminatory power is quite similar to the decision tree efficiency analysis as 92.04 percent of all units are inefficient with mean score 0.4458 and median score 0.3164. Furthermore, the kernel density functions of the scores are almost identical for the two categorizations as illustrated in Figure \ref{fig:density}.

Finally, we fit the regression model and arrive at the same conclusion -- the population remain significant while the distance is not significant. The model explains 8.34 percent of the variance of the efficiency scores which is slightly higher number than in the decision tree model. This means that the decision tree model captures environmental effects better but the two models are quite comparable.

\begin{table}
\begin{center}
\caption{Mean efficiency scores within each expert category.}
\label{tab:effExp}
\begin{tabular}{lllrrrr}
\toprule
Cat. & Population & Distance & Units & Preliminary & Dec. Tree & Expert \\
\midrule
E01 & $[   \num{0},  \num{200})$ & $(  0.00,  15.00)$ & 270 & 0.1364 & 0.4274 & 0.4181 \\
E02 & $[   \num{0},  \num{200})$ & $[ 15.00, \infty)$ & 376 & 0.1094 & 0.3369 & 0.3275 \\
E03 & $[ \num{200},  \num{500})$ & $(  0.00,  15.00)$ & 785 & 0.1155 & 0.3513 & 0.3089 \\
E04 & $[ \num{200},  \num{500})$ & $[ 15.00, \infty)$ & 682 & 0.1144 & 0.3135 & 0.3001 \\
E05 & $[ \num{500}, \num{1000})$ & $(  0.00,  15.00)$ & 741 & 0.1461 & 0.3833 & 0.3346 \\
E06 & $[ \num{500}, \num{1000})$ & $[ 15.00, \infty)$ & 474 & 0.1543 & 0.3882 & 0.4641 \\
E07 & $[\num{1000}, \num{2000})$ & $(  0.00,  15.00)$ & 463 & 0.2129 & 0.5681 & 0.5887 \\
E08 & $[\num{1000}, \num{2000})$ & $[ 15.00, \infty)$ & 249 & 0.2032 & 0.5695 & 0.6862 \\
E09 & $[\num{2000},     \infty)$ & $(  0.00,  15.00)$ & 281 & 0.4312 & 0.6546 & 0.6350 \\
E10 & $[\num{2000},     \infty)$ & $[ 15.00, \infty)$ & 201 & 0.4185 & 0.5999 & 0.8787 \\
E11 & $[   \num{0},     \infty)$ & $  0.00          $ & 138 & 0.8012 & 0.9256 & 0.9256 \\
\midrule
\multicolumn{3}{l}{All} & \num{4660} & 0.1916 & 0.4371 & 0.4458 \\
\bottomrule
\end{tabular}
\end{center}
\end{table}

\subsection{Comparison of Efficiency Scores}
\label{sec:empComp}

The preliminary efficiency analysis does not account for heterogeneous environment and we therefore do not recommend to use its efficiency scores to rank libraries. Efficiency analysis with either decision tree categories or expert categories considers environmental effects of population with town distance and is suitable to rank libraries. The categories given by the decision tree better remove the influence of the operating environment. Both approaches are, however, rather similar as the correlation coefficient between their efficiency scores is 0.8405. Preliminary efficiency scores are more different as their correlation coefficient is 0.7523 for decision tree scores and 0.7609 for expert scores.

\section{Conclusion}
\label{sec:con}

We present a comprehensive framework for the evaluation of efficiency of municipal libraries. Our study is of particular interest to public administrators involved in management of public libraries. The main finding is that there are significant differences in efficiency of municipal libraries with respect to the size of the municipality and its position in the settlement structure. A direct implication for the practice follows -- a municipal library should be compared only to the libraries with the same characteristics of the municipality.

Specifically, we assess technical efficiencies of \num{4660} public libraries established by municipalities in the Czech Republic in the year 2017. In the first stage, we adopt the Chebyshev distance DEA and utilize its many attractive properties including the super-efficiency and natural normalization. We consider total expenditures, employees and book collection as inputs with registrations, book circulation, event attendance and collection additions as outputs. In the second stage, we perform regression analysis and find that the efficiency scores are significantly dependent on the population of the municipality and distance to the municipality with extended powers. To remove the influence of the operating environment, we employ DEA for libraries separated into categories given by decision tree analysis. Interestingly, the effect of population is not completely removed suggesting it is partially environmental variable and partially explanatory variable. We also consider categories designed by an expert and find that the proposed separation approach is robust to the specification of categories to a certain degree. The proposed methodology can be used in similar applications when the data sample is large and the operating environment exhibits heterogeneity.

The main limitation of our study lies in the analyzed dataset. Although it is very extensive in terms of the number of units, it offers only few characteristics of libraries. With the available variables, we are quite capable to capture the traditional role of libraries -- making books available to the public. We are also able to assess the social and cultural role of libraries to some extent. However, we are unable to evaluate the performance of libraries in terms of the audio-visual materials, internet access and professional assistance. A more thorough efficiency analysis is needed to address all aspects of the modern library.

\section*{Acknowledgements}
\label{sec:acknow}

The author would like to thank Jan Kubát for his help with data preparation and Bojka Hamerníková, Vladimír Beneš and Marek Jetmar for their comments.

\section*{Funding}
\label{sec:fund}

The work on this paper was supported by the Technology Agency of the Czech Republic under Grant TL01000463 in the Eta program.


\end{document}